\documentclass[journal]{IEEEtran}
\usepackage{cite}
\usepackage{amsmath,amssymb,amsfonts}
\usepackage{algorithmic}
\usepackage{graphicx}
\usepackage{makecell}
\usepackage{textcomp}
\usepackage{subcaption}
\usepackage{multirow}
\usepackage{booktabs}
\usepackage{multirow}
\usepackage{hyperref}

\def\BibTeX{{\rm B\kern-.05em{\sc i\kern-.025em b}\kern-.08em
    T\kern-.1667em\lower.7ex\hbox{E}\kern-.125emX}}
\begin{document}
%\title{Detection of COVID-19 Infection on Computed Tomography Images using a Deep Learning Ensemble}
\title{COVID-19 Detection in Computed Tomography Images with 2D and 3D Approaches}
\author{Sara Atito Ali Ahmed, Mehmet Can Yavuz, Mehmet Umut Sen,\\
Fatih Gulsen, M.D., Onur Tutar, M.D., 
Bora Korkmazer, M.D., Cesur Samancı, M.D., Sabri Şirolu, M.D., Rauf Hamid, M.D., Ali Ergun Eryürekli, M.D., Toghrul Mammadov, M.D., Berrin Yanikoglu

\thanks{
%Submitted on February 4, 2021.
This work was supported by The Scientific and Technological Research Council of Turkey (TÜBİTAK) with project number 120E165.}
%1001-KAGM grant (120E165). }
\thanks{Sara Atito Ali Ahmed and Mehmet Can Yavuz are PhD candidates at Sabanci University, Istanbul, Turkey 34956; they have contributed equally to this work as first authors.
Sara is currently on research leave at University of Surrey, London, UK.
M. Umut Sen is with Urbanstat. Part of this work was done when he was a PhD student at Sabanci University (e-mails: \{saraatito,mehmetyavuz,umutsen\}@sabanciuniv.edu).}
\thanks{Fatih Gulsen, Onur Tutar, Bora Korkmazer, Cesur Samancı, Sabri Şirolu, Rauf Hamid, Ali Ergun Eryürekli and Toghrul Mammadov are with the Istanbul University-Cerrahpaşa, Cerrahpaşa Faculty of Medicine  (e-mail: 
\{fatih.gulsen,
onur.tutar,
bora.korkmazer,
cesur.samanci,
sabri.sirolu,
rauf.hamid,
ali.eryurekli,
toghrul.mammadov\}@istanbul.edu.tr }
\thanks{Berrin Yanikoglu is with the Computer Science and Engineering program at Sabanci University, Istanbul, Turkey 34956. She is also the Director of the Sabanci University Center of Excellence in Data Analytics (e-mail: berrin@sabanciuniv.edu).}

}

\maketitle

\begin{abstract}
Detecting COVID-19 in computed tomography (CT) or radiography images has been proposed as a supplement to the definitive RT-PCR test. We present a deep learning ensemble for detecting COVID-19 infection, combining slice-based (2D) and volume-based (3D) approaches. 
The 2D system detects the infection on each CT slice independently, combining them to obtain the patient-level decision via different methods (averaging and long-short term memory networks). 
The 3D system takes the whole CT volume to arrive to the patient-level decision in one step.
A new high resolution chest CT scan dataset, called the {IST-C} dataset, is also collected in this work.

The proposed ensemble, called {IST-CovNet}, obtains 90.80\% accuracy and 0.95 AUC score overall on the IST-C dataset in detecting COVID-19 among normal controls and other types of lung pathologies; and 93.69\% accuracy and 0.99 AUC score
on the publicly available MosMed dataset that consists of COVID-19 scans and normal controls only..
The system is deployed at Istanbul University Cerrahpa\c{s}a School of Medicine.

\end{abstract}

\iffalse
Detecting COVID-19 in computed tomography (CT) or radiography images has been proposed as a supplement to the definitive RT-PCR test. We present a deep learning ensemble for detecting COVID-19 infection, combining slice-based (2D) and volume-based (3D) approaches. The 2D system detects the infection on each CT slice independently, combining them to obtain the patient-level decision via different methods (averaging and long-short term memory networks). The 3D system takes the whole CT volume to arrive to the patient-level decision in one step. A new high resolution chest CT scan dataset, called the IST-C dataset, is also collected in this work. The proposed ensemble, called IST-CovNet, obtains 90.80\% accuracy and 0.95 AUC score overall on the IST-C dataset in detecting COVID-19 among normal controls and other types of lung pathologies; and 93.69\% accuracy and 0.99 AUC score on the publicly available MosMed dataset that consists of COVID-19 scans and normal controls only. The system is deployed at Istanbul University Cerrahpasa School of Medicine.
\fi

\section{Introduction}
\label{sec:introduction}
\IEEEPARstart{C}{OVID-19} is a highly contagious disease caused by the SARS-CoV-2 virus, which 
spread rapidly around the world starting early 2020 (Zhu et al. \cite{zhu2020novel}). 
The definitive diagnosis of COVID-19 is based on real-time reverse transcriptase polymerase chain reaction (RT-PCR) positivity for the presence of coronavirus \cite{corman2020detection,rubin2020role}.

Due to the long duration to obtain the RT-PCR results and the prevalence of false negative results \cite{long2020clinical}, the medical community has been in search of alternative or supplementary methods, including screening chest X-ray or Computed Tomography (CT) scans of patients for patterns of pneumonia caused by the COVID-19 infection. 

The chest X-ray  consists of a single 2-dimensional, frontal image of the thorax; consequently detection of COVID-19 infection in a chest X-ray presents as a  typical image classification problem.
On the other hand, a chest CT scan consists of a variable number of 2-dimensional axial slices, resulting in a more
challenging problem. Specifically,  the number of slices in the volume vary (typically [200-500]) and the shape and size of lung tissue within the slice vary significantly between slices.

Detecting COVID-19 in computed tomography or X-ray images has been addressed in many studies since the beginning of the pandemic
\cite{wang2020covid,narin2020automatic,hammoudi2020deep,xu2020deep,li2020artificial,wang_DeCovNet,liu2020fast}. 
Some of these systems only address the 2-class problem: distinguishing between normal and COVID-19 infected parenchyma, while others 
aim to detect COVID-19 infection among all possible conditions (normal lung parenchyma and other lung pathologies, including other types of pneumonia). The latter, which is the problem addressed in this work, is a significantly more difficult problem as non-COVID-19 pneumonia presents similar patterns to COVID-19. 
%In the remainder of this article, we will shortly refer to the simpler problem se two problems as "COVID-19 vs Healthy 

%********************************************
\begin{figure*}[t]
\centering
\includegraphics[width=0.9\linewidth]{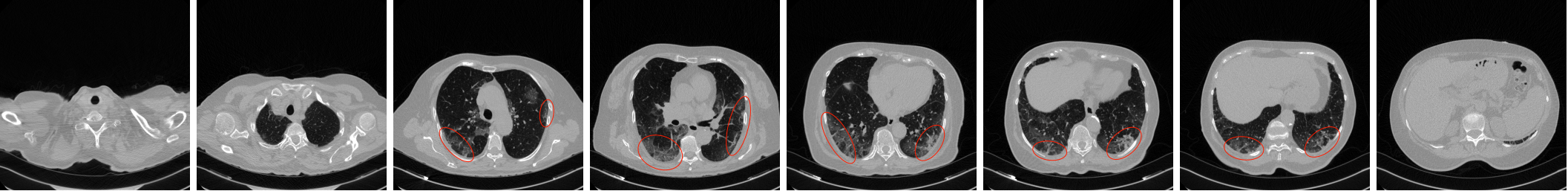}
\caption*{(a) COVID-19}
\vspace{0.2cm}

\includegraphics[width=0.9\linewidth]{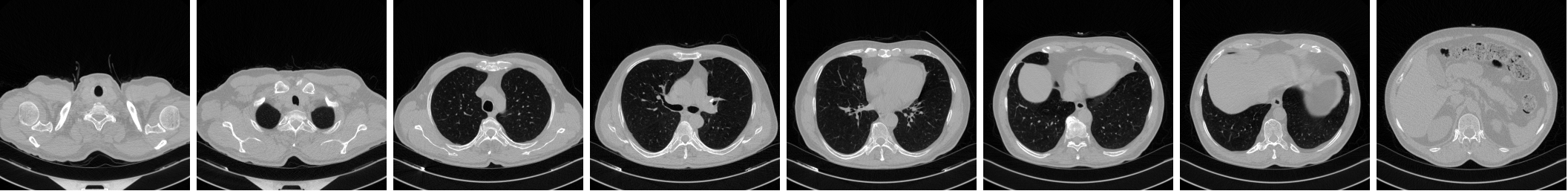}
\caption*{(b) Normal lung parenchyma}
\vspace{0.2cm}

\includegraphics[width=0.9\linewidth]{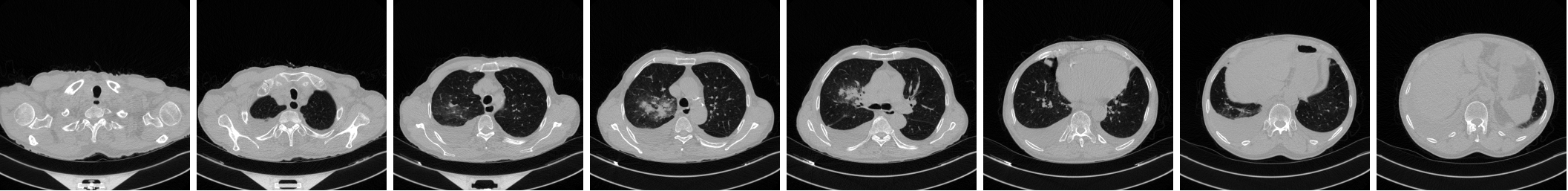}
\caption*{(c) Others  (including Non-COVID-19 pneumonia, tumors and emphysema.)}
\vspace{0.2cm}

\caption{IST-C  dataset samples. The ground glass opacities can be observed in the COVID-19 images, marked with the ellipses.  }
\label{fig:DatasetSamples}
\end{figure*}

We propose a deep learning ensemble (IST-CovNet) for detecting COVID-19 infections 
in high resolution chest CT scans, where we combine 
\textit{slice-based} and \textit{volume-based} approaches. 
The slice-based approach takes individual slices as input and outputs the COVID-19 probability for that slice. To obtain the patient-level decision from slice-level predictions, we have evaluated different classifier combination techniques, including simple averaging and Long-Short Term Memory (LSTM) networks. This system is based on transfer learning using the Inception-ResNet-V2 \cite{szegedy2017inception} network that is expended with a novel attention mechanism \cite{dang2020detection}. 
The volume-based approach is based on the {DeCoVNet} architecture of Wang et al. \cite{wang_DeCovNet} with slight modifications to the architecture.
In both approaches, we use the pretrained {U-Net} \cite{ronneberger2015u} architecture to find the lung regions in the  slice images.
Focusing to lung areas by masking the input with the lung mask is found to be an important step to reduce overfitting with such high-dimensional data \cite{gupta2020deep}.  
To obtain the patient-level decision from slice-level predictions, we have evaluated different classifier combination techniques, including simple averaging and Long-Short Term Memory (LSTM) networks. 
To combine 2D and 3D systems, we used ensemble averaging, multi-variate regression and Support Vector Machines (SVMs).

%The proposed ensemble is trained and tested using the new IST-C dataset. 
A new dataset (IST-C) is collected  at Istanbul University-Cerrahpa\c{s}a, Cerrahpa\c{s}a Faculty of Medicine (IUC),  consisting of 712 chest CT scans collected from 645 patients. 
It includes samples from COVID-19 infected patients, as well as  normal lung parenchyma and Non-COVID-19 pneumonia, tumors and emphysema patients. 
%The dataset consists of 336 chest scans from COVID-19 infected patients, along with 245 scans showing normal lung parenchyma and 131 scans from Non-COVID-19 pneumonia, tumors and emphysema patients. 
%
Figure \ref{fig:DatasetSamples} shows three samples from the IST-C dataset collected in this work, including a typical COVID-19 involvement pattern termed as \textit{ground glass opacity}, along with normal lung parenchyma and  other conditions including non-COVID-19 pneumonia, tumors and emphysema.
%
%among healthy or other lung pathologies. 
%In addition to the comparison of different classifier combination techniques, we evaluate different levels of lung tissue segmentation in  tomography images. 

The contributions of this work are the following:
\begin{itemize}
 \item We have collected 712 high resolution chest CT scans from 645 patients, showing normal lung parenchyma, COVID-19 infections, and other pathologies, including non-COVID-19 pneumonia, tumors and emphysema. 
 The IST-C dataset is made public along with our results as benchmark\footnote{http://github.com/suverim}.
 \item We present a deep neural network ensemble (IST-CovNet) that combines slice-based and volume-based approaches and achieves state-of-art accuracies on the publicly available MosMed and IST-C datasets. 
 \item We compare the two commonly used approaches in COVID-19 detection systems along with  relevant preprocessing, segmentation and combination alternatives, 
 as well as proposing novel attention and combination strategies for the slice-level approach.
 \item The system is deployed at Istanbul University Cerrahpaşa School of Medicine, to alert attending physicians for CT scans that show COVID-19 infections.

 %- can we add anything here 
 %We compare a slice-based and a volume-based approaches by evaluating them on two public datasets, together with  relevant preprocessing, segmentation and combination alternatives.
\end{itemize}

%%%%%%%%%%%%%%%%%%%%%%%%%%%%%%%%%%%%%%%%%%%%%%%%%%%%%%%%%%%%%%%%
\begin{table*}[t]
\centering
%\resizebox{\columnwidth}{!}{%
\begin{tabular}{|l|m{3.5cm}|c|c|c|c|c|c|}
\hline
\multicolumn{1}{|c|}{\textbf{Dataset}} & \textbf{Description} & \textbf{Resolution} & \textbf{\# CT Scans} & \textbf{\# Slices} & \textbf{\# COVID-19} & \textbf{\# Normal} & \textbf{\# Others}\\ \hline
%\multicolumn{8}{|c|}{2D Datasets}\\ \hline
%by Zhao et al.
%{COVID-CT \cite{zhao2020covid}}   & CT images extracted from PDF documents & Mixed & 338 & 1,305 & 224 & 114 & 0\\ \hline
%Got the stats. from https://arxiv.org/pdf/2003.13865.pdf, table 1, 2, 3, 4

%by Kumar et al.
{CC-19  \cite{kumar2020blockchain}} & CT scans collected  from 3 different hospitals  and 6 different scanners & High & 89 & 34,006 & 68 & 21 & 0 \\ \hline
%
%
%
%\multicolumn{8}{|c|}{3D Datasets}\\ \hline
%by Morozov et al.
{MosMed  \cite{morozov2020mosmeddata}}    & CT scans with indicated COVID-19 severity level (4 levels) & High & 1,110 & 46,411 & 856 & 254 & 0\\ \hline
BIMCV-COVID19
\cite{iglesia2020} & COVID-19 and Normal only & High & 	2,068 &	314,056 & 1,141	& 927	& 0\\ \hline
COVID-CT-MD \cite{afshar2020}	& COVID-19, Normal, Other & High &	305 &	45,471 & 170	& 77	& 61 \\ \hline
%stats are in end of page 7 of the pub.
%
HKBU-HPML-COVID-19 \cite{he2020}	& COVID-19, Normal, Other. Collected from different hospitals	& High &	6,878 &	406,449	& 2,513 &	1,927	& 2,435 \\ \hline
{{IST-C (this work)}}    & CT scans from one hospital & High & 712 & 200,647 & 336 & 245 & 131\\ \hline
\end{tabular}
%}
\caption{Some of the publicly available COVID-19 CT scan datasets. The first four datasets  contain scans of only COVID-19 infected patients and those with normal lung parenchyma. 
{IST-C} dataset collected in this work includes non-COVID-19 pneumonia, tumors and emphysema as well.}
\label{tbl:AvailableDatasets}
\end{table*}

\begin{table*}[t]
\centering
%\resizebox{\columnwidth}{!}{%
\begin{tabular}{|l|c|c|c|c|}
\hline
\multicolumn{1}{|c|}{} & \textbf{\# Patients} & \textbf{\# CT volumes} & \textbf{Total \# slices} & \textbf{Avg \# slices/person} \\ \hline
\textbf{COVID-19} & 300 & 336 & 92,905 & 276 $\pm$ 83 \\ \hline
\textbf{''Normal''}   & 245 & 245 & 67,712 & 277 $\pm$ 67 \\ \hline
\textbf{''Other''}    & 131 & 131 & 40,030 & 306 $\pm$ 98 \\ \hline
\textbf{Overall}    & 645 & 712 & 200,647 & 282 $\pm$ 82 \\ \hline
\end{tabular}
%}
\caption{Overview of the {IST-C} dataset: COVID-19 infections are all people diagnosed with the infection; ''Normal'' is everyone with no infection whatsoever; ''Other'' is all other types, including pneumonia,  tumors and emphysema.}
\label{tbl:DatasetStats}
\end{table*}

\section{Related Works}
\label{sec:RelatedWorks}

Automatic COVID-19 detection research in literature have targeted both chest X-rays \cite{wang2020covid,narin2020automatic,hammoudi2020deep} and CT scans \cite{xu2020deep,li2020artificial,wang_DeCovNet,liu2020fast} as input and there have been many systems published in peer-reviewed venues or pre-print sites since the beginning of the pandemic.

Comprehensive literature reviews can be found in surveys about artificial intelligence (AI) based approaches to COVID-19 in \cite{shi_general,Milon_Novel,ozsahin_survey}. 
Among these surveys, Ozsahin et al. \cite{ozsahin_survey} structure their survey into 3 groups: systems aiming to differentiate between i) COVID-19 versus normal lung parenchyma, ii) COVID-19 versus  non-COVID-19 (sometimes called COVID-19 negative) consisting of both normal lung parenchyma and other types of pneumonia, and iii) COVID-19 versus other types of pneumonia. 
Systems included in this survey report the accuracy and/or the {Area Under the Curve (AUC)} score related to the {Receiver Operating Characteristic (ROC)} curve.
State-of-art results are above 90\% accuracy and 0.95 AUC for the first problem (i) and  approximately 
88\% accuracy and 0.90 AUC for the second problem (ii). 

AI based COVID-19 detection approaches are two-fold: \textit{2D} or \textit{slice-based} approach, taking  a single slice image as input and obtain a score for individual slices \cite{narin2020automatic}, while \textit{3D} or \textit{volume-based} approach, taking the whole volume (sequence of slices) as the input and produce a single score for the patient \cite{xu2020deep,li2020artificial,wang_DeCovNet,hammoudi2020deep}.
In slice-based models, output scores of slices are often combined by averaging, to obtain the patient-level scores and decisions. 
Among volume-based approaches, most systems use adaptive-pooling operation for combining slice level features to obtain a patient-level decision \cite{li2020artificial,wang_DeCovNet}, while others use a more implicit combination using Recurrent Neural Networks (RNN) \cite{hammoudi2020deep}. 
An advantage of 2D models is the direct interpretability  while the 3D models 
is potentially more powerful as they leverage end-to-end optimization rather than a 2-stage process of obtaining patient score after slice level scores.

%%%%%%%%%%%%%%%%%%%%%%%%%%%%%%%%%%%%%%%%%%%%%%%%%%%%%%%%%%%%%
In the remainder of this section, we focus on a subset of the literature due to space limitations, reporting systems that analyze CT scans (not X-rays), address the problem of separating COVID-19 samples from all non-COVID-19 samples (not just normal lung parenchyma), and appear on peer-reviewed venues. 

%Among systems that are most relevant to our work, 
Li et. al \cite{li2020artificial} developed a model called {COVNet}, that is based on the Resnet \cite{szegedy2017inception} backbone. The varying number of CT slices are input into  parallel branches that use shared weights and the deep features extracted from each are combined by a max-pooling operation.
%and fed into a fully connected layer to generate a probability score at patient-level.
They report 0.96 AUC score on the 3-class classification problem of distinguishing between  normal lung parenchyma, COVID-19 and other lung pathologies. 
%In \cite{li2020artificial}, the datasets were collected from six hospitals between August 2016 and February 2020 and test set is independently come from the same hospitals. The ROC AUC achieved 0.96.

Wang et. al. \cite{wang_DeCovNet} use the pretrained {U-Net} \cite{ronneberger2015u} architecture to segment lung regions and obtain the lung mask volume. Then, the proposed {DeCovNet} takes the whole CT volume along with the corresponding lung mask volume as input, and outputs a patient-level probability for COVID-19. The variable number of slices is handled  using adaptive {maxpool} operation. 
Authors report \%0.91 accuracy and a 0.959 AUC score on the 2-class problem of separating COVID-19 positive cases from all others (non-COVID-19, including other pneumonia). 
%Wang et al. \cite{wang_DeCovNet} report  0.959 AUC in Covid19 vs. Non-Covid19 (including other infections), on a similar dataset of 540 patients/CT scans dataset \cite{wang_DeCovNet}. 

Hammoudi et al. \cite{hammoudi2020deep} split a chest X-ray into patches and after obtaining patch-level predictions using deep convolutional networks, they use bidirectional recurrent networks to combine them to predict patient health status.

Liu et. al \cite{liu2020fast} fine-tune well-known deep neural networks for the primary task of detecting COVID-19 and the auxiliary task of identifying the different types of COVID-19 patterns (e.g. ground glass opacities, crazy paving appearance, air bronchograms) observed in the slice-image.
%interlobular septal thickening. 
They report that using the auxiliary task helps with the detection performance, which reaches 89.0\% accuracy.

Harmon et al. \cite{harmon2020artificial} test the performance of a baseline deep neural network approach in a multi-center study. The approach consists of lung segmentation using  AH-Net \cite{ahnet} and the classification of segmented 3D lung regions by pretrained  DenseNet121 \cite{huang2017densely}. On a 1,337-patient test set they report an accuracy of 0.908 and AUC score of 0.949.

Among systems that report on the MosMed dataset, 
Jin et al. \cite{jin2020development} propose a deep learning slice-based approach employing ResNet-152 \cite{he2016deep} architecture. The developed model achieved comparable performance to experienced radiologists with an AUC score of 0.93. 
%sensitivity of 0.94, and specificity of 0.66. 
%xx Yet, the spatial correlation of the CT-scans is not considered in this work, which may decrease the overall precision of the developed approach. 
%xx what do you mean - did they do averaging??
%S: I just found this (I will send in the email) 

{He at al. \cite{he2021automated} proposed a differentiable neural architecture search framework for 3D chest CT-scans classification with the Gumbel-Softmax technique \cite{jang2016categorical} to improve the searching efficiency. The experimental results show that their automatically searched model 
%(CovidNet3D) 
outperforms three of the state-of-the-art 3D models 
%(ResNet3D101, DenseNet3D121, and MC3-18) 
achieving an accuracy of 82.29\% on MosMed dataset.}

\begin{figure*}[t!]
\centering
\includegraphics[width=0.7\textwidth]{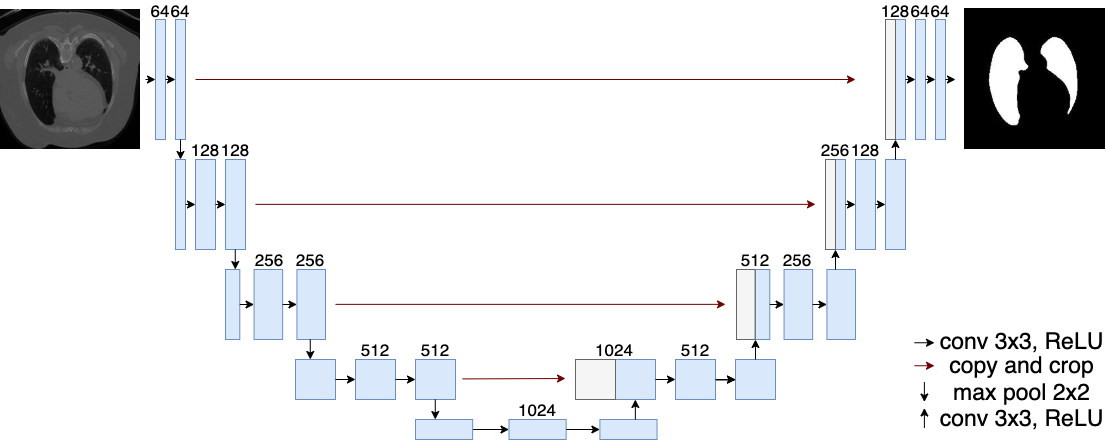}
\caption{Segmentation network {U-Net}\cite{ronneberger2015u}: input is a slice image and the output is the corresponding lung mask.}
\label{fig:unet}
\centering
\end{figure*}

%%%%%%%%%%%%%%%%%%%%%%%%%%%%%%%%%%%%%%%%%%%%%%%%%%%
%\section{Materials and Methods}
\section{{IST-C} Dataset}
\label{sec:dataset}

While there are many works on automatic detection of COVID-19 infection on X-ray or CT images, there are only a handful publicly accessible COVID-19 datasets. 
CT scan datasets we found in literature at the time of preparing this manuscript are shown in Table \ref{tbl:AvailableDatasets}. 
Note that three of these datasets, CC-19 \cite{kumar2020blockchain}, MosMed dataset \cite{morozov2020mosmeddata} and BIMCV-COVID19
\cite{iglesia2020}, only contain  COVID-19 and normal lung parenchyma. On the other hand, in MosMed, the COVID-19 samples are also labelled with the severity of the infection in 4 levels (CT-1 to CT-4).
%COVID-CT \cite{zhao2020covid} and CC-19 \cite{kumar2020blockchain} consist of images scanned from PDF documents, which have a low resolution. 
%The third one, the MosMed dataset \cite{morozov2020mosmeddata} contains only COVID-19 and normal lung parenchyma; but the COVID-19 samples are labelled with the severity of the infection in 4 levels (CT-1 to CT-4).

Lack of publicly available datasets results in researchers  collecting and reporting on their own datasets, thereby rendering a comparison between different approaches difficult.
To address this issue, we have collected a new open-source dataset called {IST-C}, retrospectively from patients admitted to the Radiology department of Cerrahpa\c{s}a Faculty of Medicine from March 2020 to August 2020. 
The collected dataset consists of 336 chest CT scans 
from COVID-19 infected patients, along with 245 scans showing normal lung parenchyma and 131 scans from Non-COVID-19 pneumonia, tumors and emphysema patients. 
These two last groups will be called simply as ''Normal'' and ''Other'' from here on. The detailed statistics of the dataset are shown in Table \ref{tbl:DatasetStats}.
%------------------------------------------------

The collected CT scans in DICOM format consists of 16-bit gray scale images of size $512 \times 512$. Each scan is accompanied with a set of personal attributes, such as patient ID, age, gender, location, date, etc. (not used in  this work).
The average age of the patients is $52 \pm 17$ years, in which $405$ of the patients are male and $274$ patients are female. 

The annotation of this dataset is at CT scan level: the CT of a patient as a whole is labelled as COVID-19, ''Normal'', or ''Other'' by expert radiologists at Istanbul University-Cerrahpaşa, Cerrahpaşa Faculty of Medicine. In the remainder of this article, we refer to {patient-level} instead of {CT-level} decisions, even though some patients may have more than one CT.

Sample images extracted from COVID-19, ''Normal'' and ''Other'' classes are shown in Figure \ref{fig:DatasetSamples}. 
The anonymized dataset is now shared publicly  at http://github.com/suverim.

%%%%%%%%%%%%%%%%%%%%%%%%%%%%%%%%%%%%%%%%%%%%%%%%%%%
\section{Preprocessing}
Pixel values of images in the CT dataset are in Hounsfield Unit (HU) which is a radiodensity measurement scale that maps distilled water to $0$ and air to $-1000$. The HU values range between $-1024$ 
and $4096$, with higher values being obtained 
from bones and metal implants in the body and lung regions typically ranging in $[-1024,0]$.

Similar to literature, we process chest CT scans such that values higher than $u_{max}=600$ are mapped to $u_{max}$ and the range $[-1024,u_{max}]$ is normalized to the $[0,1]$ linearly. 
%We used $u_{max}=600$, but we have also widened the range to 
%for the slice-based system, %but the volume-based system had slightly improved performance when we used  
%$u_{max}=2000$, to better capture the patterns of some calcifications in the  ''Other'' class (tumors, modules etc.) that appear with higher HU values.

Slice images that are originally  $(512 \times 512)$ are resized to match the input size of the  respective deep networks, namely $299 \times 299$ for slice-based system and $256 \times 256$ for the volume-based system.
For the 3D approach, we have also reduced the slice count by half, so that the whole CT volume consisting of up to around 500 slice images fits in the GPU memory. 
%We compared interpolation of two subsequent slices and skipping every other slide and found that the latter results in higher accuracy, even though interpolation is  commonly used in many biomedical applications, including segmentation \cite{johannes2020automatic}. 
This reduction is done for only the IST-C dataset where the number of slices per CT scan is high (Table II).
%, seems to have underperformed in the case of this fine  which fails in classification task.

%------------------------------------------
\section{Lung Segmentation}
\label{sec:unet}

Lung shapes vary greatly within a chest CT scan, as can be seen in Figure 1. 
%In fact, lungs  do not even appear in many of the slices in a chest CT. 
With the aim of focusing on the lung areas, we used the pretrained {U-Net} network to segment
lung regions from non-lung areas. 

The {U-Net} architecture was first proposed by Ronneberger et al \cite{ronneberger2015u} for biomedical image segmentation in general and trained specifically for lungs by Johannes et al. \cite{johannes2020automatic}. Since then has been used in detecting  lung regions  extensively in the diagnosis of lung health \cite{xu2020deep,li2020artificial,wang_DeCovNet}. 
The {U-Net} network, shown in Figure \ref{fig:unet}, is named after the U-shape formed by the encoder branch consisting of convolutional layers and the decoder branch consisting of deconvolution operations. 
The network also has skip connections in each layer, carrying the output of earlier layers to later layers. 
%The input is automatically down-sampled to  $192 \times 192$ pixels, and the output is automatically up-sampled to the input dimensions. The resize process also includes cropping to square dimensions. 

\begin{centering}
\begin{figure}[thb]
    \centering
    \includegraphics[width=0.9\linewidth]{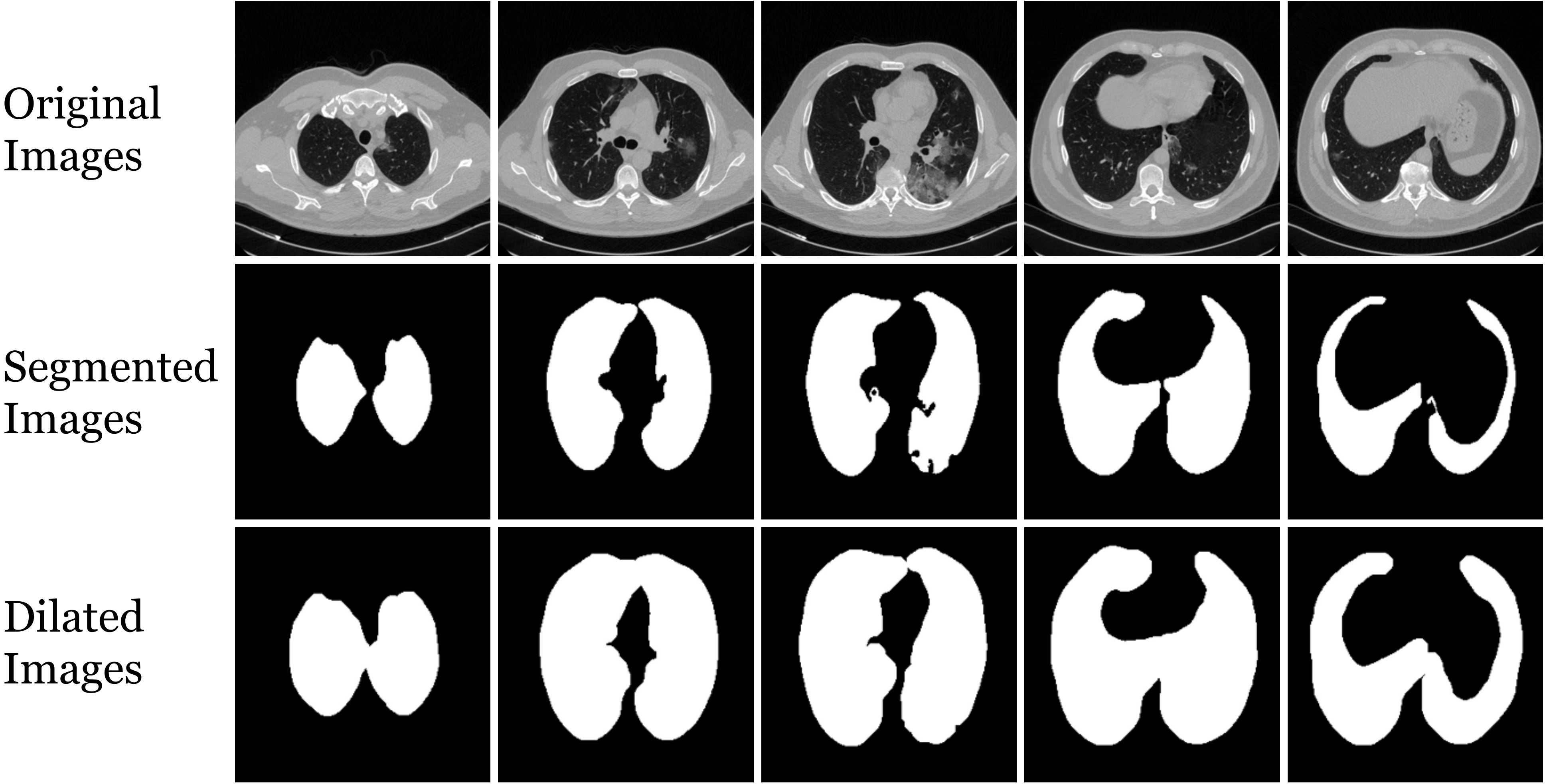}
    \caption{Sample slice  images  along  with  their  segmentation masks as obtained by U-Net and dilated masks.}
    \label{fig:SegmentedImages}
\end{figure}
\end{centering}

Lung segmentation is applied to individual slices in the CT volume. The output for each slice is the corresponding binary segmentation mask, separating lung areas (including air pockets, tumors and effusions in lung regions) from background or other organs, as shown in Figure \ref{fig:SegmentedImages}. The segmentation extracts left and right lungs separately, although this information is not used in our model. 

Lung segmentation with U-Net is very successful, as reported in \cite{johannes2020automatic} and also observed in our case. Nonetheless, in order not to miss infected regions, we 
%used mathematical morphology to fill-in the gaps in the lung masks, by using the dilation operation 
dilated the masks with a 10-pixel structuring disk.
Sample slices from the IST-C dataset and corresponding lung masks obtained by U-Net and the dilated masks are shown in Figure \ref{fig:SegmentedImages}. 
%We discuss the segmentation performance in Section \ref{sec:Exp}.

%-----2D main ----------------------------------------------------------
\begin{figure*}[th]
    \centering
    \includegraphics[width=0.85\linewidth]{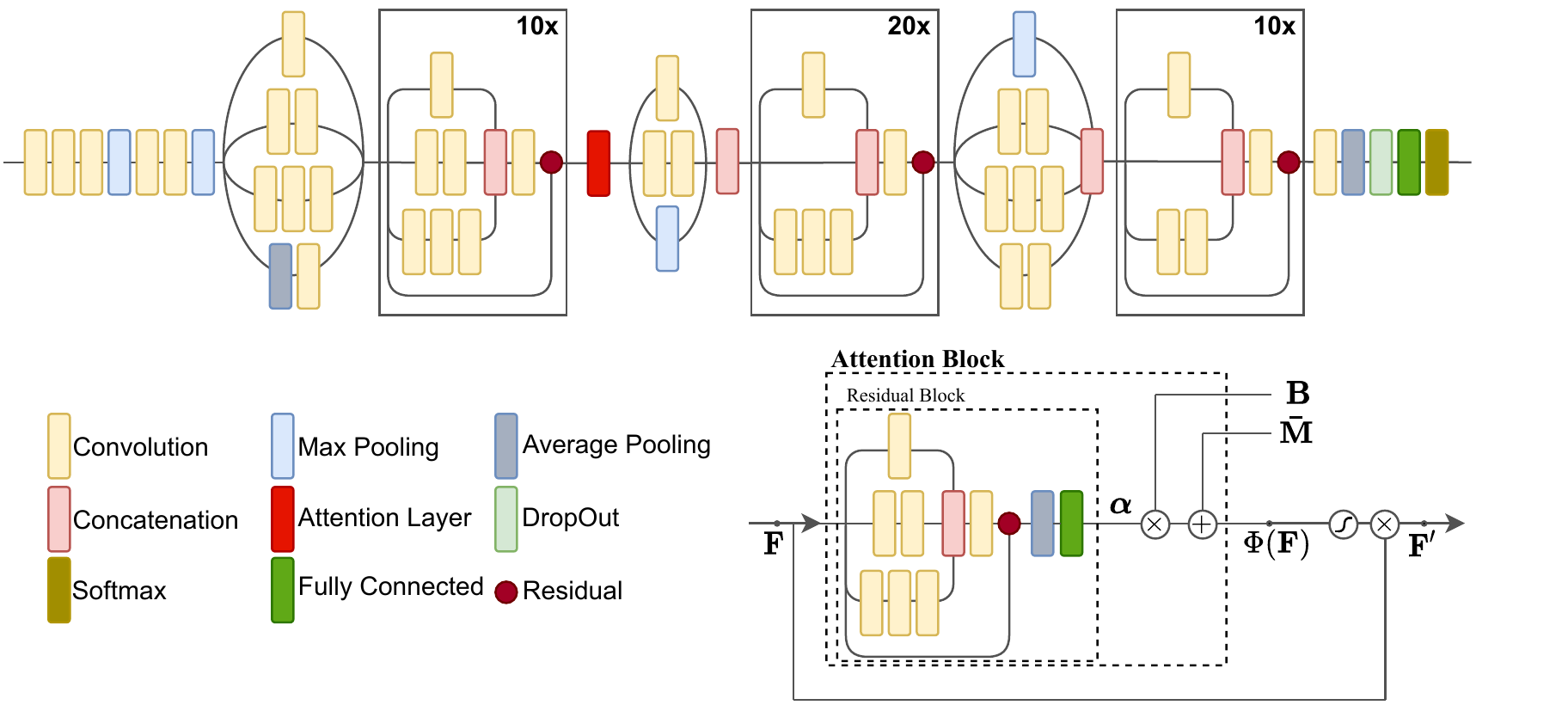}
    \caption{The base network and the inserted attention-based layer. 
    Attention layer takes the feature maps $\mathbf{F}$ as an input and estimate the attention map $\Phi(\mathbf{F})$, which is then used to attend to the original features after a sigmoid activation. }
    %Alpha is bold now in the figure
    \label{fig:AttentionLayer}
\end{figure*}

%%%%%%%%%%%%%%%%%%%%%%%%%%%%%%%%%%%%%%%%%%%%%%%%%%%
\section{Slice-based Approach}
\label{sec:SliceApproach}
In this approach, CT slices are analyzed independently, before combining them to obtain patient-level predictions. 

\subsection{Base Model}
\label{sec:baseModel}

To construct the base network architecture, 
%for the slice-based approach, 
we employed Inception-ResNet-V2 architecture \cite{szegedy2017inception}, one of the top-ranked architectures of the ImageNet Large-Scale Visual Recognition Challenge (ILSVRC) 2014 \cite{ILSVRC15}. The network architecture was used successfully in various image classification and object detection tasks \cite{ahmed2020skin,lee2019multi}.

Inception-ResNet-V2 network is an advanced convolutional neural network that combines the inception module with ResNet \cite{he2016deep} to increase the efficiency of the network. The network is $164$ layers deep with only $55.9$ million parameters.
%The image input size of the network is $299$-by-$299$.
%
It consists of three main reduction modules with $10$, $20$, and $10$ inception blocks for each module, respectively. 
The size of the output feature maps of the three reduction modules are $35\times35$, $17\times17$, and $8\times8$, respectively.

Training a large deep learning network from scratch is time consuming and  requires a tremendous  amount  of  training  data. Therefore, our approach is based on fine-tuning a pre-trained Inception-ResNet-V2 model, that is originally trained on the ImageNet dataset with 1.2 million hand-labeled images of 1,000 different object classes.

%-------------------------------------------------
\subsection{Attention Mechanism}
\label{sec:attention}

To investigate the predictions of the trained base model, we applied Class Activation Mapping (CAM) \cite{zhou2016learning} on some of the images from the validation set. Observing that the attention of the network is not always directed to the area of interest (lung tissues) in misclassified images, we decided to 
use attention maps and thereby guide the network to the
regions that are important to the problem at hand. 
%Attention mechanism has been successfully applied in many computer vision tasks, including fine-grained image recognition \cite{zheng2017learning}, face attributes classification \cite{aly2018multi}, image captioning \cite{xu2015show}, person re-identification \cite{zhao2017deeply}, face manipulation detection \cite{dang2020detection}.
Attention mechanism has been successfully applied in many computer vision tasks, including fine-grained image recognition \cite{zheng2017learning} and face attributes classification \cite{aly2018multi}.

We add an attention map block inserted to the backbone of our base network, as shown in Figure \ref{fig:AttentionLayer}. 
The input to the attention layer is a convolutional feature map $\mathbf{F} \in R^{H \times W \times C}$, where $H$, $W$, and $C$ are the height, width, and the number of channels, respectively. The output of the attention module is the masked feature map $\mathbf{F^{\prime}}
%\in R^{H \times W \times C} 
= \mathbf{F} \odot \sigma( \Phi ( \mathbf{F} ))$, obtained via element-wise multiplication of the feature maps $\mathbf{F}$ and 
sigmoid ($\sigma$) attenuated attention layer output, $\Phi ( \mathbf{F} ) \in R^{H \times W}$.

Unlike the standard approach of learning the attention layer fully within the network, the approach used
in this work is suggested to be an explainable and modular approach \cite{dang2020detection}. It makes the assumption that an attention map can be represented using the linear combination of a set of basis
vectors, as: $$\Phi ( \mathbf{F} ) = \mathbf{\bar{M}} + \mathbf{B} \times \boldsymbol{\alpha}$$
where $\mathbf{\bar{M}} \in R^{H \times W}$ is the average segmentation map; $H$ and $W$ are the height and width of the images; $\mathbf{B} \in R^{H \times W \times n}$ is the matrix of the $n$ basis vectors; and $\boldsymbol{\alpha} \in R^{n \times 1}$ are the coefficients. 

\begin{figure}[t]
    \centering
    \begin{subfigure}[t]{0.15\linewidth}
    \vspace{-1.9cm}
    \includegraphics[width=\linewidth]{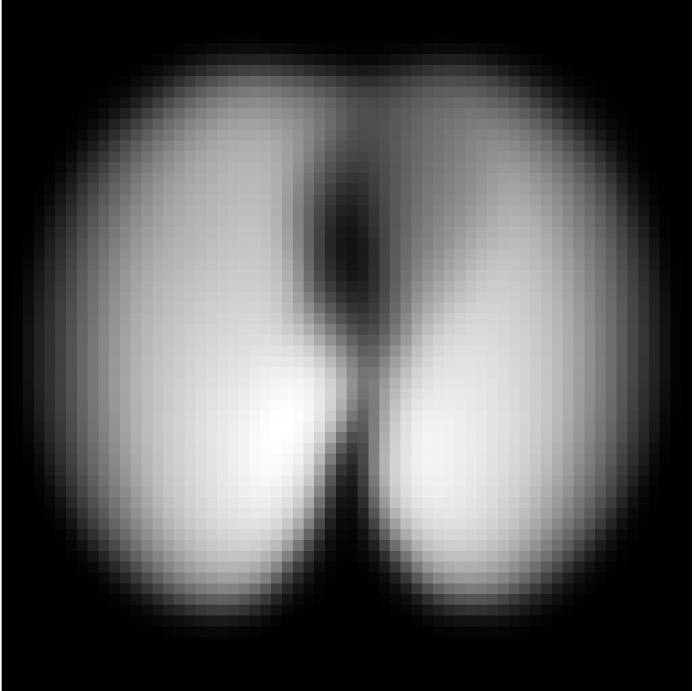}
    \caption{}
    \end{subfigure}
    \hfill
    \begin{subfigure}[t]{0.83\linewidth}
    \includegraphics[width=\linewidth]{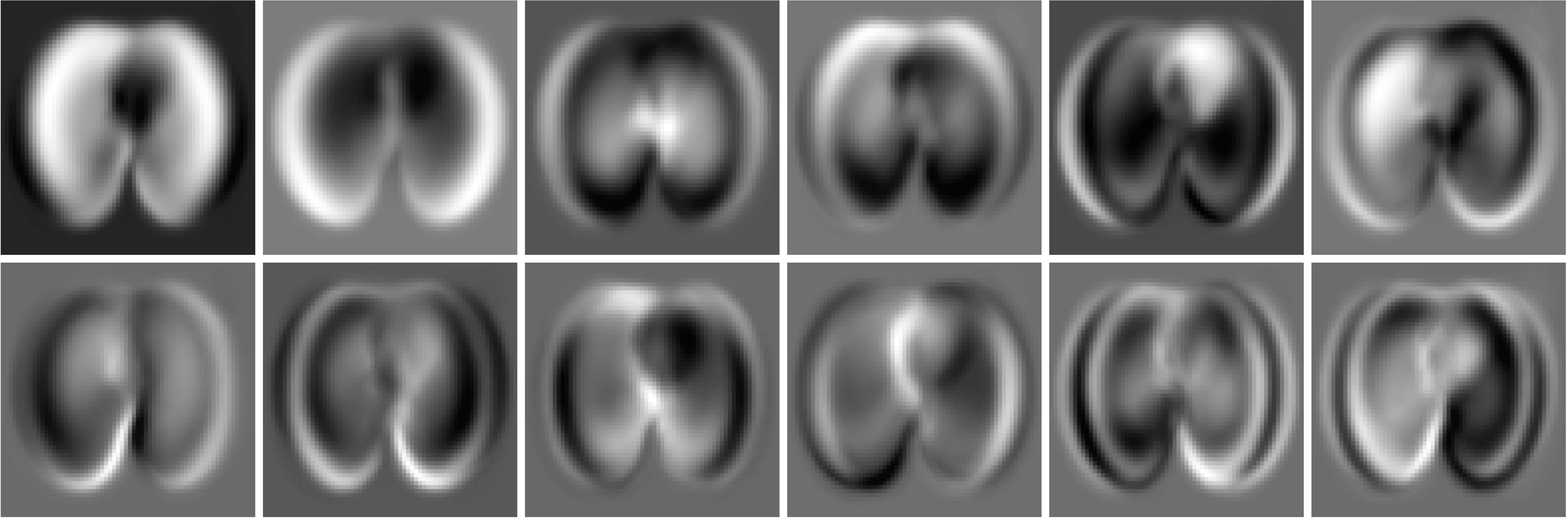}
    \caption{}
    \end{subfigure}
    \caption{(a) Mean mask $\bar{M}$ and (b) The first 12 eigenvectors.}
    \label{fig:mean_bases}
\end{figure}

The average lung map $\mathbf{\bar{M}}$ and the 12 basis vectors $\mathbf{B}$ are 
obtained by applying Principal Component Analysis to lung masks obtained by {U-Net} segmentation network. The 12 basis vectors that retain approximately 75\% of the variance are shown in Figure \ref{fig:mean_bases} and U-Net is 
explained in Section \ref{sec:unet}.

To obtain the attention map coefficients $\boldsymbol{\alpha}$
%$\Phi ( F )$,
an additional convolutional block is inserted to the network getting the input from the feature maps $\mathbf{F}$, as shown in Figure \ref{fig:AttentionLayer}. 
The convolutional block consists of a separable convolutional layer
%followed by a batch normalization layer and a ReLU activation function. The separable convolutional layer 
which is a depth-wise convolution performed independently over each
channel of an input, followed by a pointwise convolution, batch normalization, and ReLU activation function.
The output of the convolutional block 
(or attention coefficient block) are the weights $\mathbf{\alpha}$ which form the coefficients in the linear basis vector representation.

%--------------------------------------------
\subsection{Implementation Details }

The Inception-ResNet-V2 network used as the base model in the slice-based approach is chosen due to its relatively small size and good performance.
The network has an RGB image input size of $299\times299$. The output layer of the model is replaced with a fully connected layer with $2$ hidden units to represent the given classes: COVID-19 vs Non-COVID-19 (including ``Normal`` and ``Other`` samples). 
% !this is not a binary classification network, binary has 1 output with 0 or 1, this has 2 outputs and the loss is cross entropy function.
All the layers in the classification network are finetuned and optimized using categorical cross-entropy loss function.

For the attention based model, we added the attention layer after the first reduction block as shown in Figure \ref{fig:AttentionLayer}. As for the attention loss function, we trained the network in unsupervised manner. Even in the absence of the attention map supervision, we found that the attention module is able to learn the discriminative regions automatically.

The implementation is done using  the  Inception-ResNet-V2 model provided in the \mbox{Matlab} deep learning toolbox. Besides, several commonly used data augmentation techniques are applied during training, such as rotation [$-5$ to $5$], $x$ and $y$ translation [$-5$, $5$], and $x$ and $y$ scaling [$0.9$, $1.1$].

Throughout this work, we set the batch size equal to $64$ and the initial learning rate as 1e-5 with a total of $50$ epochs with \mbox{Adam} algorithm for parameters optimization. The training process takes around $100$ minutes per epoch for IST-C dataset and $40$ minutes per epoch for MosMed datasets employing 8GB Nvidia GeForce RTX 2080 GPU.

%--------------------------------------------
\subsection{Combining Slice-level Predictions}
\label{sec:stop}
The straightforward approach to obtain patient-level decision is to combine the predictions of the slice based model using simple averaging of slice-level predictions. This is evaluated as the base model, to obtain the patient-level score. 

However, simple averaging does not take into account the information about the characteristics of COVID-19 infection, such as the fact that the patterns are often seen in the lower parts of the lungs.
To learn this type of information about the slice sequence and to also handle the variable length of the slice sequence,  
we also used Recurrent Neural Networks (RNNs) as an alternative \cite{rumelhart1986learning}.

We used Long Short-Term Memory (LSTM) network \cite{hochreiter1997long} that is the most powerful type of recurrent network.
%, to recognize patterns in sequences of data as they have a temporal dimension that allows them to take time and sequence into account. 
The input to the network consists of deep features corresponding to each slice in the CT volume. The features are extracted from the last pooling layer of the slice-based CNN model with the attention module, 
discussed in Section \ref{sec:SliceApproach}). The LSTM learns to combine the slice-level features  to obtain patient-level predictions.

The LSTM architecture consists of 3 layers:  i) a bidirectional LSTM layer with $1024$ hidden units and a dropout layer to reduce overfitting; ii) another bidirectional LSTM layer with $512$ hidden units;
 and iii) a fully connected layer with an output size corresponding to the number of classes (2 or 3 in our case).

It is important to note that the number of slices in the CT volumes varies substantially which can introduce lots of padding into the training process of the LSTMs and consequently  negatively impact the classification accuracy. To overcome this issue, we normalized each CT sequence into 282 slices (i.e. the mean slice count across the IST-C dataset), by either dropping or replicating slices depending on the length of the volume. 
After normalization, each slice of the CT volume is passed to the trained CNN model for feature extraction. Then, the LSTM model is trained using the sequence of the feature vectors corresponding to the slices.

%Using LSTMs compared to simple averaging resulted in \%1-2 points improvement over simple averaging, as discussed in Section \ref{sec:Exp}.
%%%%%%%%%%%%%%%%%%%%%%%%%%%%%%%%%%%%%%%%%%%%%%%%%%%%%%%%%%

%%%%%%%%%%%%%%%%%%%%%%%%%%%%%%%%%%%%%%%%%%%%%%
\section{Volume-base Approach}
\label{sec:3DApproach}

The 3D volume-based approach takes as input the whole CT volume and outputs  patient-level decision,  based on a single step processing of the input. It uses the lung segmentation volume obtained by U-Net (described in \ref{sec:unet}), followed by a classification network based on DeCoVNet \cite{wang_DeCovNet}.

The segmentation network ({U-Net}) takes as input a single slice of the chest CT and outputs a binary mask indicating the lung region. The classification network 
subsequently takes the CT volume and the corresponding binary mask volume and outputs the patient-level scores.

%---------------------------------------------------
\subsection{Classification network}
\label{sec:classification}
The classification network used in our work is based on {DeCoVNet} that has been proposed by Wang et al. \cite{wang_DeCovNet}. We have made some modifications to this network, without significantly changing its architecture. 
%Nonetheless, we will call this modified version 3D-C19, for clarity.

\begin{figure*}[t!]
\centering
\includegraphics[width=0.8\textwidth]{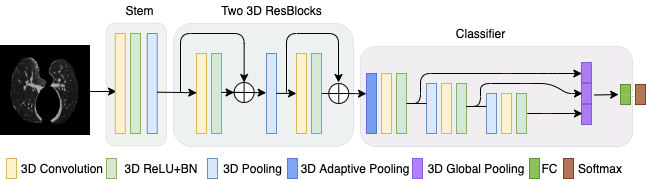}
\caption{Architecture of the classification network which is based on {DeCoVNet} \cite{wang_DeCovNet}. }
\label{fig:3dnetwork}
\centering
\end{figure*}

The network consists of three consecutive blocks, (1) Stem (2) ResBlocks (3) Classifier, as shown in Figure \ref{fig:3dnetwork} and detailed in Table \ref{DC19}.
The stem block consists of a convolutional layer with a receptive field size \scalebox{.9}[1.0]{5x7x7} (depth, height, width), as used in well-known networks AlexNet \cite{krizhevsky2012imagenet} and Resnet \cite{he2016deep}. The convolutional layer is followed by a batchnorm layer and a pooling layer. 
We evaluated using both a single channel input, consisting of the slice image with the lung mask applied, 
as well as the 2-channel input, consisting of the input slice  and its lung mask, as in the original network. 
As we expected, the 2-channel approach led to less efficient training and did not bring accuracy gains.
%We observed that the 2-channel input led to less efficient training.  

%---------------------------------------------------
The second stage of the networks consists of two 3D residual blocks (ResBlocks), with maxpool operation in between to reduce the volume depth by half \scalebox{.9}[1.0]{64xT/2x64x64}. In each  block, there are 2 kernels: \scalebox{.9}[1.0]{3x1x1}, \scalebox{.9}[1.0]{1x3x3} (depth,height,width) with a stride of 1 in each dimension and padding of 1 wherever needed. The output volume is of size \scalebox{.9}[1.0]{128xT/2x32x32}. This block is adopted  without any modification.

The third block, called the Progressive classifier, starts with an adaptive maxpool operation that handles the variable number of slices and outputs \scalebox{.9}[1.0]{128x16} feature maps of size $32 \times 32$. It is followed  by 3 convolution layers and pooling operations, followed by a fully connected output layer with softmax activation. 

The main modification in this block is to enrich the  feature representation. 
The original DeCoVNet had a global max pooling layer with $32 \times 1 \times 1 \times 1$ nodes, in the penultimate layer.
We extended the Progressive classifier block by adding a new layer of concatenated features obtained using global max pool operation after each of the 3D convolutional layers. 
More specifically, from a convolutional layer with \scalebox{.9}[1.0]{FxDxHxW} output volume, the global max pooling operation outputs a vector of size $F$. 
The resulting  192-dimensional \scalebox{.9}[1.0]{($96+48+48$)}  feature vector is fully connected to the output layer (2 nodes with softmax activation), as shown in Figure \ref{fig:3dnetwork}. 
We thus increased the penultimate layer size from 32 to 192. 

This feature representation was inspired by the work in \cite{ahmed2019within}, where authors proposed to  approximate a deep learning ensemble by replicating the output layer with connections from earlier layers and extending the loss function to include all the loss terms \cite{ahmed2019within}.
The  classification network architecture is given in Table \ref{DC19}.

\begin{table}[thb]
\centering
\resizebox{\columnwidth}{!}{%
\begin{tabular}{llll}
           & Operation                   & Output   & Penult. \\ 
\hline
Stem       & Conv3d@5x7x7  & 16xTx64x64        &               \\ 
\hline
ResBlocks  & ResBlock@3x1x1\&1x3x3  & 64xTx64x64       &                  \\
           & MaxPool3d                   & 64xT/2x64x64    &                   \\
           & ResBlock@3x1x1\&1x3x3 & 128xT/2x64x64        &              \\ 
\hline
Progressive  & AdaptiveMaxPool3d           & 128×16×32×32     &                  \\
Classifier           & Conv3d@3×3×3  & 96×16×32×32              &          \\   
           & GlobalPool3d   &  &  96x1x1x1           \\ 
           & ------------ 2nd Block& ------------  & \\ 
           & MaxPool3d           & 96×4×16×16            &           \\
           & Conv3d@3×3×3  & 48×4×16×16                       \\
           & Dropout3d (p=0.5)            & 48×4×16×16                        \\ 
           & GlobalPool3d  &            &   48×1×1×1           \\
           & ------------ 3rd Block&------------ & \\ 
           & MaxPool3d           & 48×4×16×16      &                 \\
           & Conv3d@3×3×3+ReLU   & 48×4×16×16          &              \\    
           & GlobalPool3d   &   &    48x1x1x1           \\
           & FullyConnected     & 2       &        \\
\hline
\end{tabular}
}
\caption{The 3D-classification network  architecture. The residual blocks have two kernels.}
\label{DC19}
\end{table}

%%%%%%%%%%%%%%%%%%%%%%%%%%%%%%%%%%%%%%%%%%%%%%%%%%%%%%%%
\subsection{Implementation Details}
 
We train the network in two stages: first with the 1,110-sample MosMed dataset that  contains only COVID-19 and healthy classes, as pretraining. 
%This step is important so as to increase the size of the data used to learn the network weights.
%
In the second stage of training, we fine-tuned the network on the IST-C dataset, which contains samples from other pneumonia conditions as well. 

In both  stages, the settings are the same and as follows: the loss function is the categorical cross-entropy; the  optimizer is the Adam optimizer used with 1e-5 learning rate. Since the graphical card Nvidia 2080 can only process a single batch at a time, the batch size is one due to memory constraints. We also used data augmentation exactly same with DeCoVNet: scaling ($1-1.2$), rotation ($10$ degrees) and translation ($0-10$  pixels). 

{All 3D systems were run for 200 epochs and validation set accuracy was observed. 
%Even if the training loss nearly zero, the training continued because there is augmentation. 
The optimal weights were chosen as those giving the highest validation set results and applied to the test set to get probability distribution over the COVID-19/Non-COVID-19 classes.}

{The 3D systems take around 8 minutes for an epoch of IST-C dataset and 4 minutes for an epoch of MosMed dataset in 8Gb Nvidia 2080.}

\section{Combining Multiple Systems}

After training the 2D and 3D systems, we combine output of the systems (\textit{patient-level} predictions)  to obtain the final prediction. 
In contrast, please note that Section \ref{sec:stop} discusses the combination of \textit{slice-level} predictions to obtain patient-level predictions for the 2D approach.

The 2D (slice-based) approach is realized with or without the attention mechanism and using different combination mechanisms to obtain the patient-level decision. 
Similarly, the 3D (volume-based) approach is realized with 1-channel input where the input is masked with the lung mask, or with 2-channel input as in the original DeCovNet \cite{wang_DeCovNet}.

The combination methods that were evaluated were averaging, multivariate linear regression and Support Vector Machines (SVM). However, we only report ensemble averaging results because multi-variate regression essentially assigned the same weights to the two combined systems and the SVM did not bring noticeable improvements to justify the more complex combination method.
%have an accuracy to warrant the more complex combination method.

%%%%%%%%%%%%%%%%%%%%%%%%%%%%%%%%%%%%%%%%%%%%%%%%%%%%%%%%%%%%%

\section{Experimental Evaluation}
\label{sec:Exp}

We have trained and evaluated the proposed 2D and 3D approaches to COVID-19 detection, for the IST-C and MosMed datasets that are described in Section \ref{sec:dataset}. These results are given in Tables \ref{tbl:results} and \ref{tbl:Mosmed_results}.
We also report the test results of the trained ensemble on the COVID-CT-MD dataset \cite{afshar2020}, to evaluate inter-operability performance.  These results are given in Table \ref{tbl:results_COVID-CT-MD}.

We split the IST-C database into training/validation/testing data. For "COVID-19" class, $100$ volumes are used for testing and the rest are used of the training and the validation. For "Normal" and "Others" classes, $100$ and $50$ volumes are used for testing, respectively. In total, we assigned $250$ volumes for testing and $462$ for training and validation. The MosMed dataset was split randomly as train-test, with a 80-20\% split, resulting in a $222$ test samples.

We first trained both systems (2D and 3D) on MosMed training set and tested the ensemble on MosMed test set. For IST-C, the 2D system was trained using only the training portion of IST-C data set, while the 3D system was first pretrained with MosMed and finetuned on IST-C training set. 

We have done some extensive evaluation comparing different preprocessing, segmentation,  architecture and ensemble methods. 
However for the sake of clarity, we report only the most important experiments,
using accuracy and AUC scores, in line with the literature. 

%***********************************************
\begin{table}[t]
\centering
\resizebox{\columnwidth}{!}{%
\begin{tabular}{|c|c|c|}
\hline
\textbf{~~Model~~}
& \textbf{~~Accuracy (\%)~~} & \textbf{~~~AUC~~~} \\ \hline
%\multicolumn{4}{|c|}{\textbf{\begin{tabular}[c]{@{}c@{}}2-Class Problem \end{tabular}}}  \\ \hline

\multicolumn{1}{|l|}{\begin{tabular}[c]{@{}c@{}}2D - Base Network + Averaging\end{tabular}}    
& 80.80 $\pm$ 4.88 & 0.87    \\ 
\multicolumn{1}{|l|}{\begin{tabular}[c]{@{}c@{}}2D - Base + Attention + Averaging\end{tabular}} 
& 85.60 $\pm$ 4.35 & 0.90     \\ 
\multicolumn{1}{|l|}{\begin{tabular}[c]{@{}c@{}}2D - Base + Attention + LSTM \end{tabular}} 
& \textbf{87.20 $\pm$ 4.14} & \textbf{0.89}    \\ \hline
\multicolumn{1}{|l|}{3D - DeCoVNet \cite{wang_DeCovNet}} & 78.00 $\pm$ 5.14   & 0.78\\
\multicolumn{1}{|l|}{3D - two-channels } & 81.45 $\pm$ 4.82 & 0.86  \\ %skip
%\multicolumn{1}{|l|}{3D - 1-channel - interpolation } & 82.80 & 0.86  \\ 
\multicolumn{1}{|l|}{3D - one-channel }  & \textbf{87.20 $\pm$ 4.14} & \textbf{0.90} \\%skip \Xhline{1\arrayrulewidth} 
\multicolumn{1}{|l|}{Ensemble - Averaging (IST-CovNet)}
& \textbf{90.80 $\pm$ 3.58}  & \textbf{0.95} \\\hline
%\multicolumn{1}{|l|}{Ensemble - SVM } 
%& \textbf{88.00}  & \textbf{0.95} %\\\hline
\end{tabular}
}
\caption{Performance results for the IST-C test set with $n=250$ samples. The 2D systems were trained with only IST-C and the 3D systems were trained with MosMed and IST-C training subsets. Bold figures indicate the best accuracy in slice-based or volume-based approaches. }%Ensemble results were obtained with the best  2D and 3D systems. }
\label{tbl:results}
\end{table}

%\multicolumn{1}{|l|}{Sys13: Ensemble-overall - SVM w/ linear kernel (systems Sys1-Sys4))} & \textbf{91.5}  & ???  & ???  \\\hline
%\multicolumn{1}{|l|}{\begin{tabular}[l]{@{}l@{}}Regression\\ (Repeat COVID in Validation) \end{tabular}} 
%& 87.60  & 10.00 & 16.00  &  88.00 \\  \Xhline{1\arrayrulewidth}

%***********************************************
\begin{table}[t]
\centering
\resizebox{\columnwidth}{!}{%
\begin{tabular}{|c|c|c|}
\hline
\textbf{~~Model~~} & \textbf{~~Accuracy (\%)~~}   & \textbf{~~AUC~~}  \\ \hline

\multicolumn{1}{|l|}{\begin{tabular}[c]{@{}c@{}}Jin et al. (2D) \cite{jin2020development} \end{tabular}} 
& -  & 0.93 \\ 
\multicolumn{1}{|l|}{\begin{tabular}[c]{@{}c@{}}He et al. (3D) \cite{he2021automated} \end{tabular}} 
& 82.29  & -   \\ 
\multicolumn{1}{|l|}{3D - DeCoVNet \cite{wang_DeCovNet}} 
& 82.43  & 0.82   \\\hline
\multicolumn{1}{|l|}{\begin{tabular}[c]{@{}c@{}}2D - Base + Attention + Averaging \end{tabular}} 
& 90.09 $\pm$ 3.93 & 0.96    \\ 
\multicolumn{1}{|l|}{\begin{tabular}[c]{@{}c@{}}2D - Base + Attention + LSTM \end{tabular}} 
& \textbf{91.89 $\pm$ 3.59}  & \textbf{0.95}   \\ \hline
\multicolumn{1}{|l|}{3D - one-channel } 
& \textbf{93.24 $\pm$  3.30}  & \textbf{0.96}  \\ \Xhline{1\arrayrulewidth}
%as there are other works now in the table, we should specify ensemble of what. Just mentioning here until the table is complete in order not to forget
\multicolumn{1}{|l|}{Ensemble - Averaging  (IST-CovNet)}
& \textbf{93.69 $\pm$ 3.20} & \textbf{0.99} \\\hline
\end{tabular}
}
\caption{Performance results for the MosMed \cite{morozov2020mosmeddata} test set with $n=222$ samples. Our approaches were trained using only MosMed training subset. Bold figures indicate the best accuracy in slice-based or volume-based approaches.The dataset only contains COVID-19 and Healthy scans. 
%Since the CT scans have fewer slices, slice skipping was not necessary.
}
\label{tbl:Mosmed_results}
\end{table}

\begin{table}[ht]
\centering
\resizebox{\columnwidth}{!}{%
\begin{tabular}{|c|c|c|}
\hline
\textbf{~~Model~~}
& \textbf{~~Accuracy (\%)~~} & \textbf{~~~AUC~~~} \\ \hline
%\multicolumn{4}{|c|}{\textbf{\begin{tabular}[c]{@{}c@{}}2-Class Problem \end{tabular}}}  \\ \hline

\multicolumn{1}{|l|}{\begin{tabular}[c]{@{}c@{}}COVID-FACT \cite{dialameh} \end{tabular}} & 91.83 & - \\ 
\multicolumn{1}{|l|}{\begin{tabular}[c]{@{}c@{}}CT-CAPS \cite{ctcaps} \end{tabular}} & 89.80 & 0.93 \\ 
\multicolumn{1}{|l|}{\begin{tabular}[c]{@{}c@{}}Deep-CT-Net \cite{covidfact} \end{tabular}} & 86.00 & 0.886 \\ \hline
\multicolumn{1}{|l|}{Ensemble Averaging  (IST-CovNet)} & {87.86 $\pm$ 3.67}  & {0.92} \\\hline
\end{tabular}
}
\caption{Performance results for the COVID-CT-MD \cite{afshar2020} test set   with $n=305$ samples. Our ensemble system was trained using \textit{only}  MosMed and IST-C datasets. }%Ensemble results were obtained with the best  2D and 3D systems. }
\label{tbl:results_COVID-CT-MD}
\end{table}

%%%%%%%%%%%%%%%%%%%%%%%%%%%%%%%%%%%%%%%%%%%%%%%%%

\subsection{COVID-19 vs. Non-COVID-19}
\label{subsec:Res}

The results  for the 2-class problem (distinguishing COVID-19 from all Non-COVID-19 scans) are given in Tables \ref{tbl:results} and \ref{tbl:Mosmed_results} for IST-C and MosMed respectively. 

The best results obtained on the IST-C dataset is 90.40\% accuracy and 0.95 AUC score with ensemble averaging of the best 2D and and best 3D system. 
The results obtained on the MosMed dataset with only COVID-19 and Normal classes are better, given the relatively simpler problem with two classes. Our ensemble method achieved 93.69\% accuracy and 0.98 AUC score which is 10\% points higher than state-of-art, as indicated in Table \ref{tbl:Mosmed_results}. More importantly, the AUC score is 0.06 higher compared to the best AUC score \cite{Jin2020CT}.

One of the motivations of this work was to compare the effectiveness of 2D and 3D approaches. Considering the 
results given in Tables \ref{tbl:results} and \ref{tbl:Mosmed_results}, we see that the best 2D and 3D approach have the same accuracy 
on the IST-C datasets (87.20\%), while the 3D system is slightly better for the MosMed dataset (93.24\% vs 91.89\%). 

As for improvements brought by novel approaches in our systems, 
the attention layer increased the accuracy significantly (85.60\% vs 80.80\% in IST-C),
%which is in line with the results obtained in other image understanding problems. 
and the use of LSTM brings another 1-1.2\% points improvements in accuracy for both datasets, compared to averaging the slice-level predictions to obtain the patient-level prediction.

For the 3D approach, we observed that the 2-channel input also used in DeCoVNet achieves significantly lower accuracy (81.45\% vs 87.20\%), probably due to the difficulty in training the first layer weights. 
The supplied code for DeCoVNet  \cite{wang_DeCovNet} also achieved lower results compared to our modified version (82.43\%) vs  87.20\% for IST-C and 82.43\% vs 93.24\% for MosMed). 
%
%Additionally, the interpolation done to halve the CT volume in the case of the IST-C dataset leads to significantly lower  performance (\%87.20 vs \%82.80), presumably due to the loss of the fine details in the images. 

\begin{figure*}[t]
    \centering
    \begin{subfigure}[t]{0.4\linewidth}
    \includegraphics[width=\linewidth]{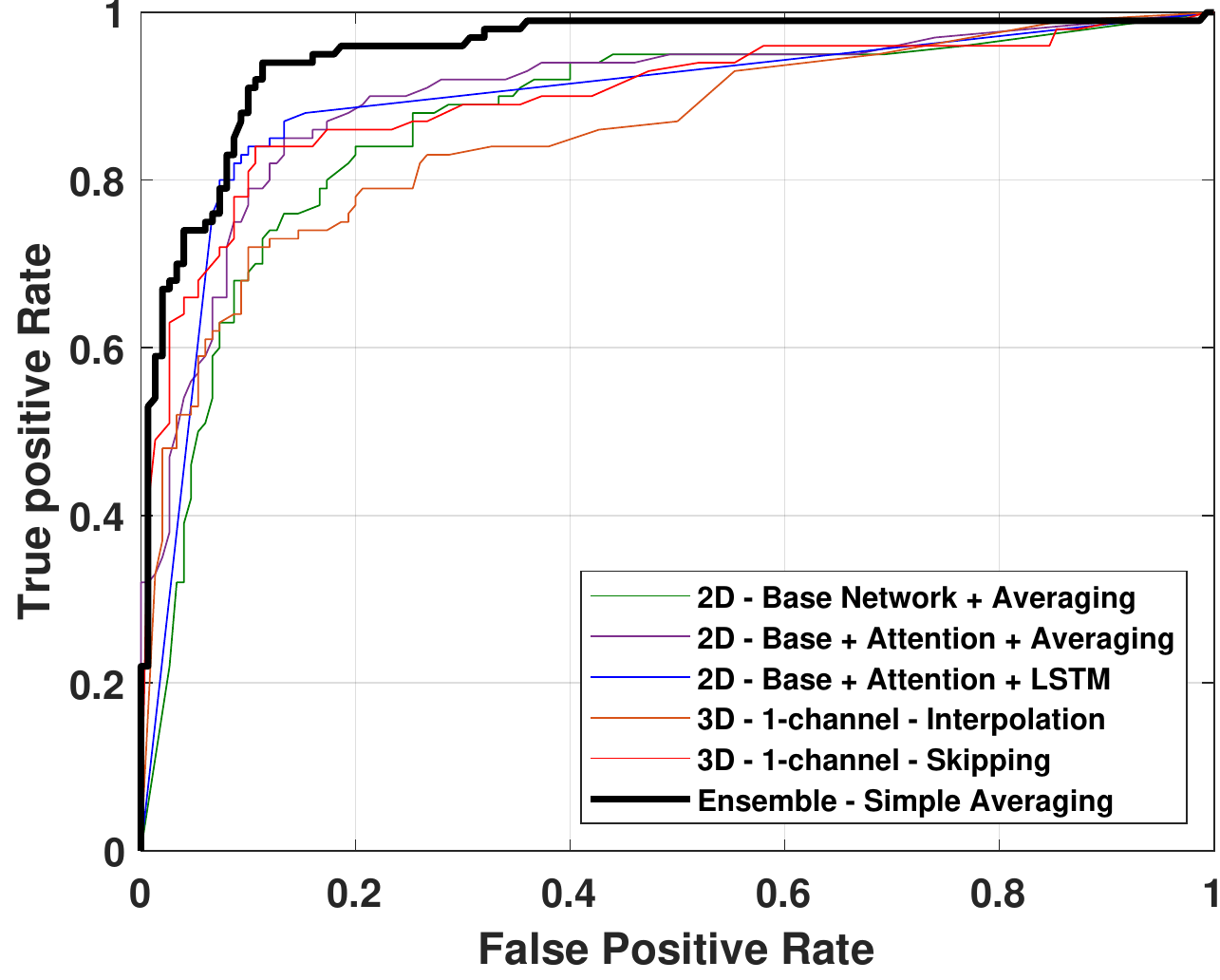}\caption{}
    \end{subfigure}
    \qquad
    \begin{subfigure}[t]{0.4\linewidth}
    \includegraphics[width=\linewidth]{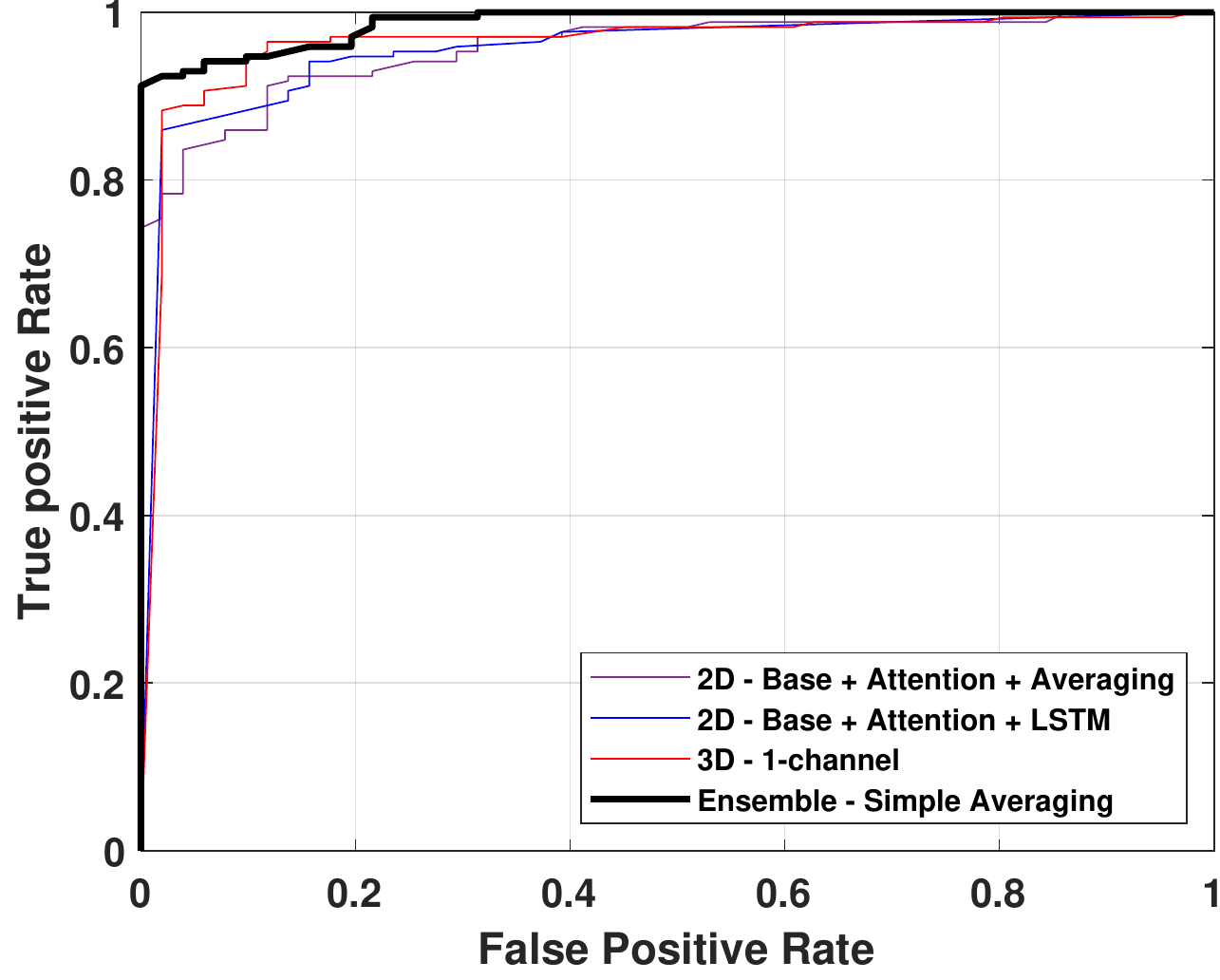}\caption{}
    \end{subfigure}
    \caption{ROC curves of the trained models on (a) IST-C dataset and (b) MosMed dataset.
    \label{fig:Roc_curves}}
\end{figure*}

The accuracy values are given together with 95\% confidence intervals that are computed using the Wilson score interval method \cite{wilson} for the number of test samples in each dataset.
ROC figures corresponding to IST-C and MosMed datasets are given in Figure \ref{fig:Roc_curves}.

%---------------------------------------------
\subsection{Inter-Operability}

To study the inter-operability of systems with respect to different tomogrophy equipment/settings, we  tested the accuracy of the system trained on MosMed and IST-C datasets on COVID-CT-MD dataset \cite{afshar2020}. The results shown in Table \ref{tbl:results_COVID-CT-MD} accuracy and AUC results (87.86\% and 0.92) are better than one of the state-of-art results on that dataset \cite{ctcaps} and indicate a small accuracy decrease compared to the IST-C dataset results (90.90\% vs 87.86\%).

\begin{figure}[t]
    \centering
    \includegraphics[width=0.95\linewidth]{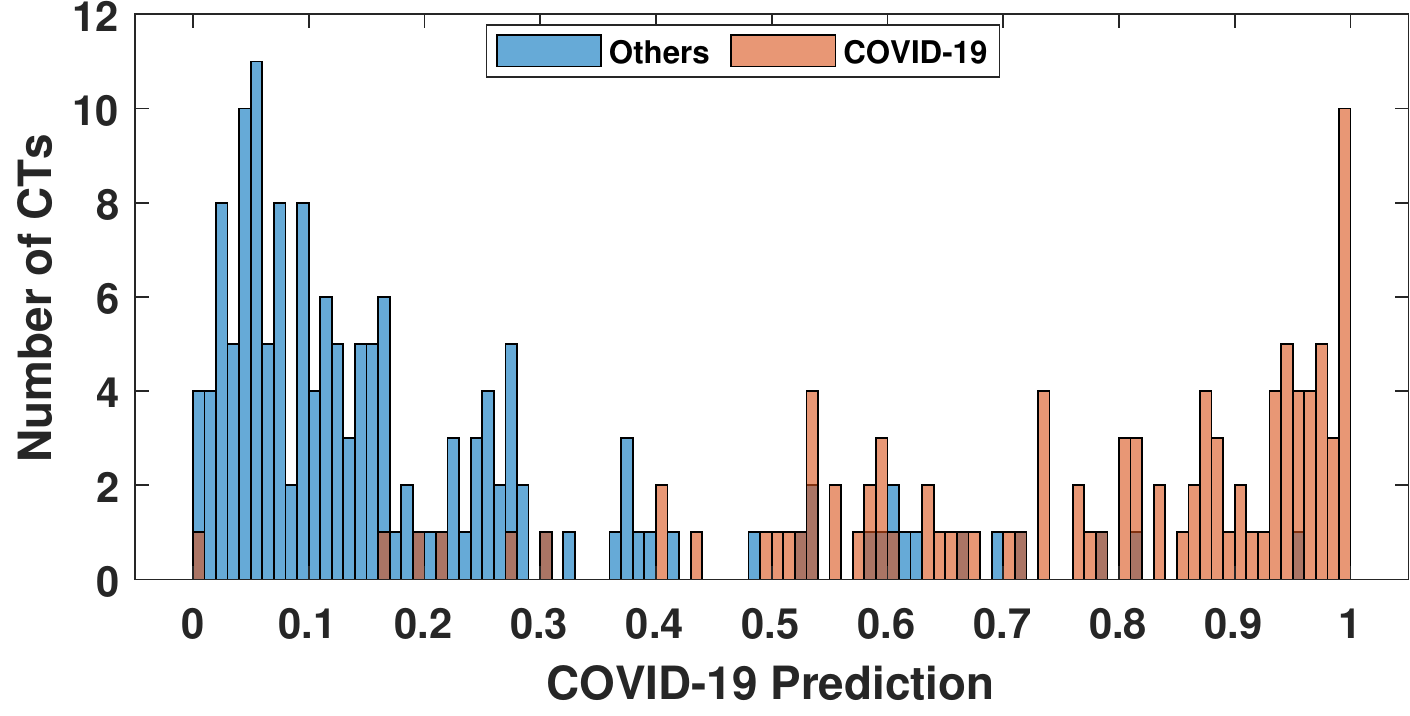}
    \caption{COVID-19 predicted probability distribution for the IST-C dataset, using the ensemble. }
    \label{fig:prob_responses}
\end{figure}

%Acc = 90.09 - AUC = 0.96 - FP = 17.65 - FN = 7.60
%Acc = 91.89 - AUC = 0.95 - FP = 15.69 - FN = 5.85
%Acc = 93.24 - AUC = 0.96 - FP = 17.65 - FN = 3.51
%Acc = 93.69 - AUC = 0.99 - FP = 21.57 - FN = 1.75

%%%%%%%%%%%%%%%%%%%%%%%%%%%%%%%%%%%%%%%%%%%%%%%%%%%%%%%%%%%%%
\subsection{Prediction Scores Distribution}

The system is designed to alert the attending physicians in case of sufficiently high COVID-19 probability. Hence, we also considered the COVID-19 prediction distribution of the ensemble, shown in 
Figure \ref{fig:prob_responses}. An adjustable threshold (e.g. 0.3-0.4) can be set to
alert the attending physician, at the risk of some increased False positives. 

At 0.3 threshold, we obtain 95.0\% sensitivity (true positive rate) and 80.0\% specificity (1-false positive rate) on the IST-C test data set. The ROC figures of the ensemble, corresponding to IST-C and MosMed datasets are given in Figure \ref{fig:Roc_curves}.

%------------------------------------------------------

%%%%%%%%%%%%%%%%%%%%%%%%%%%%%%%%%%%%%%%%%%%%%%%%%%%%%%%%%%%%%
\begin{figure*}[t!]
\centering
\includegraphics[width=\textwidth]{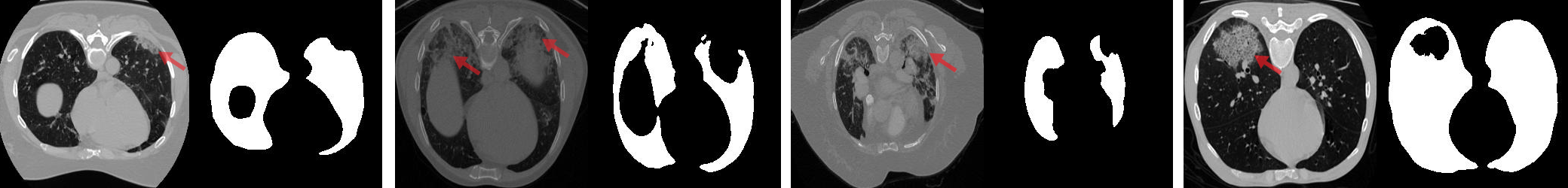}
\caption{Samples of segmentation errors (a) slice image  (b) corresponding lung masks. Problematic areas are indicated with red arrows and are often missed lung tissue due to infection or tumors.}
\label{fig:plungmask}
\centering
\end{figure*}

\subsection{Lung Segmentation Results}

Regarding lung segmentation accuracy,  Hofmanninger et al \cite{johannes2020automatic} report 97-98\% Dice similarity scores measuring how much the mask generated by U-Net and ground-truth overlaps, on different test datasets involving multiple lung pathologies. While their tested datasets also  included  ground glass opacities observed in COVID-19 cases, we evaluated the segmentation network's performance specifically for the COVID-19 detection problem
%, as it directly affects the performance of the 3D approach.
by visually checking the segmentation results of 5 slices from sampled at regular intervals from 1,156 CT scans (all covid patients from IST-C  and MosMed datasets), for a total of 5,783 slice images. We found around 11 serious segmentation errors, corresponding to roughly \%0.19, which is in line with \cite{johannes2020automatic}. 
Samples of these images are given in \ref{fig:plungmask}, where lung areas that are considered as background and are highlighted by ellipses. 
Noting that the errors occur only in some of the slices within one CT scan, we conclude that U-Net provides a successful segmentation, suitable for COVID-19 detection. 

%Note that lung masks are used to mask the slice images in the proposed 3D approach with a single channel input, in both training and testing.
%, while the slice-level  approach uses the masks in a more limited fashion, as discussed in \ref{sec:attention}. 
%In contrast, the slice-level approach uses lung segmentations only to extract the mean lung shape and  the eigen-vectors describing lung shape variations that are used in the attention block, \textit{prior} to training the classification network.

%%%%%%%%%%%%%%%%%%%%%%%%%%%%%%%%%%%%%%%%%%%%%%%%%%%%%%
\subsection{Discussion}
\label{subsec:Discuss}

While our 3D approach is based on DeCoVNet  \cite{wang_DeCovNet}, we were able to outperform its results on both datasets, thanks to the changes made 
to the model.
In particular, using only one input channel leads to more efficient training, especially since the U-Net lung segmentation is very accurate and enriching the network architecture also contributed to higher accuracy. 

Similarly, even though the 2D system is based on fine-tuning a pretrained deep network, the use of the novel attention mechanism and LSTMs to combine slice-level features bring significant improvements
over the base network and the standard approach of averaging slice predictions. 
We are aware of another work that also combines a deep network with LSTMs, related to COVID-19 predictions: Hammoudi et al. \cite{hammoudi2020deep} use bidirectional LSTMs to predict patient health status by combining the predictions made by a deep network for \textit{image-patches} of an X-ray.

%The developed system is deployed at  Istanbul University Cerrahpa\c{s}a School of Medicine in April of 2021 to flag CT scans that have a high COVID-19 score as predicted by the system. 

%The results of the slice-based system are slightly better than those of the volume-based system. While this may be counter-intuitive, training the network with the whole volume may be a more complicated task. Similar results have been observed in the Kaggle competition ``Deepfake Detection Challenge \footnote{\href{https://www.kaggle.com/c/deepfake-detection-challenge/}{https://www.kaggle.com/c/deepfake-detection-challenge/}}'' where the task is to identify videos with facial or voice manipulations. The winner of this Kaggle competition stated that the frame-by-frame classification approach performed better than the other complex approaches  \footnote{\href{https://github.com/selimsef/dfdc\_deepfake\_challenge}{https://github.com/selimsef/dfdc\_deepfake\_challenge}}.

\section{Conclusion}
%A conclusion section is not required. Although a conclusion may review the  main points of the paper, do not replicate the abstract as the conclusion. A conclusion might elaborate on the importance of the work or suggest  applications and extensions.

In addition to presenting a state-of-art system, we provide an evaluation of different 2D and 3D approaches on two datasets and discuss the effects of relevant preprocessing, segmentation and classifier combination steps on performance.

The collected dataset (IST-C) is made public to contribute to the literature as a challenging new dataset that consists of high resolution chest CT scans from a variety of conditions.

This work was motivated to help combat the pandemic and the developed system (IST-CovNet) is deployed and in use at Istanbul University Cerrahpa\c{s}a School of Medicine, to flag suspected COVID-19 cases when the patient is still at the tomography room. 
%and  speed up their medical evaluation and reducing the spread of the virus.

\nocite{*}
\bibliographystyle{IEEEtran}  
\bibliography{tmi}  

\end{document}